\documentstyle[aps,prb,epsfig]{revtex}

\begin{document}
\twocolumn[\hsize\textwidth\columnwidth\hsize\csname@twocolumnfalse\endcsname
\draft

\title{Superconductivity in the quasi-two-dimensional Hubbard model}

\author{Xin-Zhong Yan}

\address{Institute of Physics, Chinese Academy of Sciences, P.O. Box603, Beijing 100080, China\\
E-mail: yanxz@aphy.iphy.ac.cn}

\date{\today}

\maketitle

\widetext
\begin{abstract}
On the basis of spin and pairing fluctuation-exchange approximation, we study the superconductivity in quasi-two-dimensional Hubbard model. The integral equations for the Green's function are self-consistently solved by numerical calculation. Solutions for the order parameter, London penetration depth, density of states, and transition temperature are obtained. Some of the results are compared with the experiments for the cuprate high-temperature superconductors. Numerical techniques are presented in details. With these techniques, the amount of numerical computation can be greatly reduced.
\end{abstract}

\pacs{PACS numbers: 74.62.-c,74.20.-z,74.72.-h,74.40.+k}

\vfill
\narrowtext

\vskip2pc] 

\section{Introduction}

The Hubbard model has been considered as the basic model to study the mechanism of high-temperature superconductivity in the cuprates.\cite{Anderson} By this model, the spin-fluctuation-exchange between electrons is considered as responsible for the mechanism of high-temperature superconductivity. A number of calculations, taking into account of the spin-fluctuation effects, have been devoted to investigating the superconducting properties of the two-dimensional Hubbard models.\cite{Weng,Schrieffer,Vignal,Bickers,Pines,Pao,Monthoux,Dahm,Putz,Yamada,Moriya,Ueda} 

It is proved that the spin-fluctuation theory can successfully describe a number of properties, including the temperature dependences of the antiferromagnetic correlation length \cite{Dahm} and the electric resistivity,\cite{Moriya1} of the cuprates at high temperatures. However, in most of the calculations on the Hubbard model, the superconducting pairing is treated by the mean-field like approximation. Such an approximation is not appropriate because the pairing fluctuation is significant in the low-dimensional superconducting systems.\cite{Emery,Babaev,Tesanovic,Levin,Yan,Yanase} In fact, the pairing fluctuation can result in new physical consequence. It is believed that the pairing fluctuation is relevant to the pseudogap phenomena\cite{Levin,Yan,Randeria,Daggoto,Kyung,Kontani} observed in the normal state\cite{Williams,Renner} as well as in the superconducting state\cite{Tallon1} of the cuprates. 

One of the approaches treating the pairing fluctuation is the ladder-diagram approximation, which has been developed on the quasi-two-dimensional (Q2D) phenomenological model\cite{Levin,Yan} and also on the two-dimensional Hubbard model. \cite{Yanase} By the ladder-diagram approximation, the long-wavelength fluctuation is taken as the predominant contribution. It has been shown that the pairing fluctuations can result in considerable reduction of the transition temperature $T_c$. According to this approach, $T_c$ vanishes in the absence of inter-layer coupling. The reason is that the pairing fluctuation is divergently strong in the two-dimensional system. This is consistent with the Mermin-Wagner-Hohenberg (MWH) theorem.\cite{Mermin} 

In this work, we intend to study the pairing-fluctuation effect on the Q2D Hubbard model. In addition to the spin-fluctuation-exchange (S-FLEX), we take into account of the contribution from the pairing fluctuation in the self-energy of the one-particle Green's function. With this spin-pairing-fluctuation-exchange (SP-FLEX) approximation, we investigate the superconductivity in the Q2D Hubbard model. By self-consistently solving the integral equations for the Green's function, we calculate the order parameter, London penetration depth, density of states (DOS), and transition temperature. Some of the results are compared with experiments for the cuprate high-temperature superconductors. In the meanwhile, we also present some numerical techniques in details in the appendices, which is necessary for carrying out the numerical solution for the Green's function. 

\section{Formalism}

The Q2D Hubbard model defined on a layered cubic lattice is of the following form\cite{Hubbard}
\begin{equation}
H= - \sum_{\langle{\bf ij}\rangle,\>\alpha}
t_{\bf ij} c_{{\bf i}\alpha}^{\dagger}
c_{{\bf j}\alpha} + U \sum_{\bf i}
n_{{\bf i}\uparrow} n_{{\bf i}\downarrow}-\mu\sum_i(n_{{\bf i}\uparrow}+n_{{\bf i}\downarrow})
\label{Hamiltonian}
\end{equation}
where $t_{\bf ij}$ denotes the hopping energy of electron between the lattice sites $\bf i$ and $\bf j$, $c_{{\bf i}\alpha}^\dagger$ ($c_{{\bf i}\alpha}$) represents the electron creation (annihilation) operator of spin-$\alpha$ at site $\bf i$, $n_{{\bf i}\alpha}=c_{{\bf i}\alpha}^\dagger c_{{\bf i}\alpha}$, $U$ is the on-site Coulomb interaction, and $\mu$ is the chemical potential. The $\langle{\bf ij}\rangle$ sum runs over the nearest-neighbor (NN) sites. In the following, we shall assume $t_{\bf ij}=t$ for the intra-layer NN hopping and $t_{\bf ij}=t_z$ for the interlayer NN hopping. A quasi-two-dimensional system is characterized by the condition $t_z/t\ll 1$. Throughout this paper, we use unites in which $\hbar=k_B=1$.

\subsection{Normal state}

For simplifying the Feynman diagrams for the Green's function, we here firstly present the approximation scheme for the normal state. The result for the superconducting state can be immediately obtained by adding the anomalous Green's function contributions, and will be presented in the next subsection. The normal Green's function for the electrons is given by
\begin{equation}
G({\bf k},\>z_n)=\frac{1}{z_n - \xi_{\bf k} - \Sigma({\bf k},\>z_n)} \label{Green's function}
\end{equation}
where $z_n=i(2n-1)\pi T$ with $T$ the temperature of the system is the imaginary fermionic Matsubara frequency, $\xi_{\bf k}=-2t(\cos k_x+\cos k_y)-2t_z\cos k_z-\mu$, and $\Sigma({\bf k},\>z_n)$ stands for the electron self-energy. 
For brevity, occasionally, we use the generalized momentum $k = ({\bf k},z_n)$ in this paper. 

Figure 1 shows the approximation scheme for the self-energy. The first two diagrams in Fig. 1(a) are the well-known S-FLEX approximation. These two diagrams can be combined into a single diagram by redefining an effective interaction $V_{\rm eff}$ that is the summation of two interactions given by Figs. 1(b) and 1(c). The expression for $V_{\rm eff}$ is\cite{Bickers,Pao,Monthoux}
\begin{equation}
V_{\rm eff}(q)
= \frac{3}{2}\frac {U^2\chi(q)}{1+U\chi(q)}
+ \frac{1}{2}\frac{U^2\chi(q)}{1-U\chi(q)}
- U^2\chi(q)
\label{Effective Potential}
\end{equation}
with
\begin{equation}
\chi(q)=\frac{T}{N}\sum_k
G(k + q) G(k). \label{Polar Function}
\end{equation}
The generalized momentum $q$ stands for $({\bf q}, Z_m)$ with $Z_m = i2m\pi T$ the bosonic Matsubara frequency. The first and second terms in right-hand-side of Eq. (\ref{Effective Potential}) come respectively from the spin and charge fluctuations. The last term eliminates a double counting in the second-order diagrams. Owing to the predominant spin fluctuation, it is named as spin-fluctuation-exchange approximation. The Hartree term has been neglected since it is a constant, which can be absorbed in the chemical potential. In the present calculation, we add the third diagram in Fig. 1(a) representing the contribution from the pairing fluctuation. Apart from two interaction sides, the shaded part essentially represents the processes of the electron pair propagating. Figure 1(d) gives the ladder-diagram approximation for it with the second order term given by Fig. 1(e), which describes the basic propagating process. The pairing interaction between two electrons of opposite spins contains two parts, one due to the transverse spin fluctuation (TSF) as given by Fig. 1(c), and another one the screened Coulomb potential (SCP) given by Fig. 1(f). In the right-hand-side of diagrammatic equation of Fig. 1 (e), the first diagram represents the propagating of a pair without changing their spins in the intermediate state since they interact through SCP during the process. In the second diagram, the intermediate spin configuration is changed because the two electrons interact through the TSF. The third diagram describes the process the two electrons firstly interact through SCP and then through the mediation of TSF, with a minus factor stemming from once appearance of TSF. The last diagram is similar to the third one but with inverse interaction sequence. For brevity, we have dropped all the momentum on these diagrams. The momentum and spins attached to the ladder-diagram are illustrated in Fig. 2. For the sake of discussion, we here introduce a notation $L_{\alpha\beta,\beta' \alpha'}(k,q-k;q-k',k')$ for the ladder-diagram. In following, we will show that at $T \leq T_c$ the value of the ladder-diagram diverges at the long-wavelength limit, $q \to 0$. Therefore, the pairing fluctuation represented by the ladder-diagram gives significant contribution to the self-energy.

\vskip -3mm
\begin{figure}
\centerline{\epsfig{file=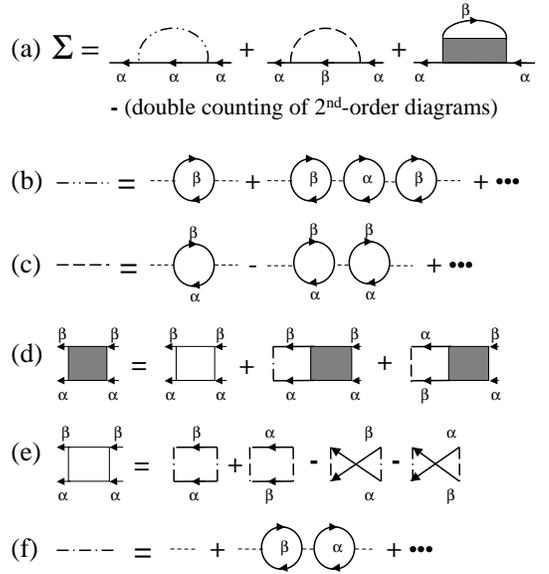,width=8 cm}}
\vskip 2mm
\caption{Approximation scheme for the self-energy. (a) Self-energy for the $\alpha$-spin electrons. The first term comes from the coupling of the $\alpha$-spin electrons with the density fluctuation of opposite $\beta$-spin electrons. The second term is due to the coupling between transverse spins through their fluctuation. The last term represents the contribution from the pairing fluctuation. (b) Interaction between $\alpha$-spin electrons due to the density fluctuation of $\beta$-spin electrons. (c) Interaction between transverse spins stemming from their fluctuation. (d) Ladder-diagram approximation to the pairing fluctuation. (e) Second order ladder-diagrams. (f) Screened Coulomb interaction between electrons of opposite spins.}
\end{figure}

To see how the pairing fluctuation takes effect, we consider Fig. 1(d) for the case of a pair of electrons with opposite spin and opposite momentum that is the case of the ladder-diagram at the long-wavelength limit. Because of $L_{\alpha\beta,\beta' \alpha'}(k,-k;-k',k')=-L_{\beta\alpha,\beta' \alpha'}(-k,k;-k',k')$, we thereby can combine the last two terms in Fig. 1(d) with an effective pairing interaction $V_{\rm P}$ defined by Fig. 3(b), and obtain an equation like to Fig. 3(c) but with `$\approx$' replaced by `=' under the ladder-diagram approximation. To solve this equation, we expand the effective pairing interaction in terms of a complete set of basis functions $\phi_n$,
\begin{equation}
V_{\rm P}(k,k')= \sum_n v_n \phi_n(k) \phi_n(k'). \label{pair interaction}
\end{equation}
The function $\phi_n$ satisfies the eigen-equation [see Fig. 3(c)]
\begin{equation}
\frac{T}{N}\sum_{k'}V_{\rm P}(k,k')G(k')G(-k')\phi_n(k')=\lambda_n\phi_n(k)  \label{eigen eq}
\end{equation}
where $N$ is the total number of lattice sites, and $\lambda_n$ is the eigenvalue. By so doing, we get
\begin{equation}
L_{\alpha\beta,\beta' \alpha'}(k,-k;-k',k')=-\sum_n\frac{\lambda_n v_n\phi_n(k)\phi_n(k')}{1-\lambda_n}.  \label{L1 eq}
\end{equation}
At the transition temperature $T_c$, the largest eigenvalue equals unity, by which the eigen-equation (\ref{eigen eq}) then reduces to the gap equation. At the superconducting state, the eigen-equation (\ref{eigen eq}) is modified by adding the term of the anomalous Green's functions with the largest eigenvalue being unity unchanged. Therefore, at $T \leq T_c$, $L_{\alpha\beta,\beta' \alpha'}(k,-k;-k',k')$ is infinitive. It implies that the long-wavelength pairing fluctuation gives significant contribution to the self-energy. On the other hand, as $T \to T_c$ from the normal state, with the largest eigenvalue of Eq. (\ref{eigen eq}) approaching to unity, $L_{\alpha\beta,\beta' \alpha'}(k,-k;-k',k')$ diverges at this limit. Therefore, for the normal state at $T$ close to $T_c$, the long-wavelength pairing fluctuation is important as well.

\begin{figure}
\centerline{\epsfig{file=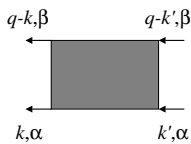,width=2.5 cm}}
\vskip -2mm
\caption{Ladder-diagram representing the propagating of a pair of total momentum $q$. $k'$ and $q-k'$ are the initial momentum of the $\alpha$ and $\beta$-spin electrons, respectively. After propagating, their momentum change to $k$ and $q-k$.}
\end{figure}

From the right-hand side of Eq. (\ref{L1 eq}), we see that except for the term corresponding to the largest eigenvalue, all other terms are finite. We therefore keep only the most diverging term to simplify the solution of the equation given by Fig. 1(d). That is, we can consider only the pairing of largest eigenvalue.\cite{Kontani} For the present case, the largest one is the $d$-wave pairing. Hereafter, we denote the largest eigenvalue and the corresponding eigen-function simply as $\lambda_d$ and $\phi(k)$, respectively, and the coupling constant simply by $v$.

\begin{figure}
\centerline{\epsfig{file=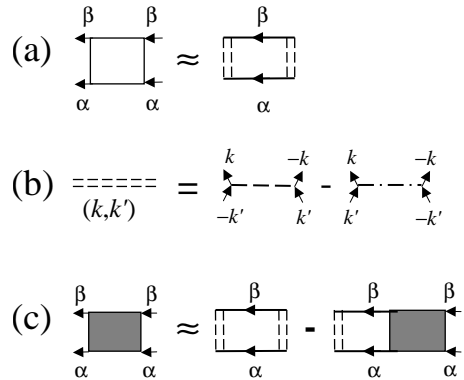,width=6 cm}}
\vskip -3mm
\caption{Approximation to the ladder-diagram. (a) Approximate second order ladder-diagram obtained from Fig. 1(e) by neglecting the dependence of the interactions on the total momentum of the pair. (b) Effective pairing interaction. (c) Renormalized diagrammatic equation for the ladder-diagram.}
\end{figure}

On observing the above-mentioned fact, we make the approximation in Fig. 1(d) using the pairing interactions of zero total momentum because near $q =0$ the pairing fluctuation is most significant. We then obtain equations as given by Fig. 3 for determining $L_{\alpha\beta,\beta' \alpha'}(k,q-k;q-k',k')$. Further more, by considering only the $d$-wave pairing that has the largest eigenvalue, the ladder-diagram given by Fig. 3(c) reduces to the same one as we previously encountered for the phenomenological model.\cite{Yan} Applying the previous result to the present case (see Appendix A), the ladder-diagram summation is obtained as 
\begin{equation}
L_{\alpha\beta,\beta \alpha}(k,q-k;q-k',k')=P(q)\phi(k)\phi(k'), \label {L eq}
\end{equation}
with
\begin{eqnarray}
P(q)&=&\frac{v^2\Pi(q)}{1+v\Pi(q)}-v^2\Pi(q) \label {P eq}\\
\Pi(q)&=&-\frac{T}{N}\sum_k\phi^2(k) G(k) G(q-k). \label{P Polar Function}
\end{eqnarray}
In Eq. (\ref{P eq}), the last term eliminates all the second-order diagrams since the contribution to the self-energy from the first two diagrams in the right-hand-side of the diagrammatic equation of Fig. 1(e) has been taken into account in S-FLEX approximation, while the other two diagrams are negligible with comparing to the infinitive ladder-diagram summation. For the self-energy, the final expression is
\begin{equation}
\Sigma(k)=-\frac{T}{N}\sum_qG(k-q)V_{\rm eff}(q)+\frac{T\phi^2(k)}{N}\sum_q G(q-k)P(q). \label {self energy}
\end{equation}
As we have noted above, the last term in Eq. (\ref{self energy}) is corresponding to the previous approximation\cite{Yan} for the phenomenological model.\cite{Levin,Yan} In that model, however, the interaction is simply a constant $d$-wave pairing potential, and the hopping energy is proportional to the hole concentration for taking into account of the constraint excluding the double occupation. The prohibition of double occupation stems from the $t-J$ model that is the large $U$ limit of the Hubbard model. It is a consequence of strong short-range antiferromagnetic coupling. Under the SP-FLEX approximation scheme for the Hubbard model, however, the antiferromagnetic coupling is taken into account by the first term in Eq. (\ref{self energy}). Moreover, for not too large $U$, the $d$-wave pairing potential given above varies with the temperature, hole concentration and $U$.
 
By passing this subsection, we give an expression for the pairing potential. As seen from Figs. 1(c), 1(f) and 3(b), $V_{\rm P}(k,k')$ equals $V_{\rm 1(c)}(k+k')-V_{\rm 1(f)}(k-k')$. Since the pairing function $\phi(k)$ can be taken as even function of $k$, the pairing potential in Eq. (\ref{eigen eq}) can be thereby written as $V_{\rm P}(k-k')$. We then have 
\begin{equation}
V_{\rm P}(q)
= \frac{3}{2}\frac {U^2\chi(q)}{1+U\chi(q)}
- \frac{1}{2}\frac{U^2\chi(q)}{1-U\chi(q)}
- U.
\label {p-poten}
\end{equation}
Therefore, the left-hand-side of Eq. (\ref{eigen eq}) is a convolution of $V_{\rm P}$ and the rest part.

In addition, the chemical potential $\mu$ should be determined to yield the hole concentration 
\begin{equation}
\delta = -\frac{T}{N}\sum_k[G(k)+G(-k)].
\label{delta}
\end{equation}
All the above equations form the closed system that self-consistently determines the Green's function.

\subsection{Superconducting state}

For the superconducting state, the above results should be extended to including the contributions from the anomalous Green's function. In the Nambu representation, the Green's function is given by
\begin{equation}
\hat{G}(k) = [z_n -\xi_k\sigma_3-\hat{\Sigma}(k)]^{-1} \label{M-Green}
\end{equation}
where $z_n$ is understood as $z_n\sigma_0$, and $\sigma$ is the Pauli matrix. Occasionally, we will use the Pauli components of $\hat G$ defined by $\hat G = G_0 + G_1\sigma_1 +G_3\sigma_3$. Correspondingly, the self-energy is expressed as $\hat \Sigma = \Sigma_0 +\Sigma_1\sigma_1 + \Sigma_3\sigma_3$ as well. The diagonal element $\Sigma_{11}(k) = \Sigma_0(k) + \Sigma_3(k)$ is given by the same diagram Fig. 1(a) except where the effective interaction and the ladder-diagram should include the contribution from the anomalous Green's function. The element $\Sigma_{22}(k)$ is obtained by $\Sigma_{22}(k) = - \Sigma_{11}(-k)$. The off-diagonal part $\Sigma_1$ is given by the gap equation
\begin{equation}
\Sigma_1(k) = -\frac{T}{N}\sum_{q}G_1(k-q)V_{\rm P}(q).  \label{gap-eq}
\end{equation}
The expressions for $V_{\rm eff}$ and $V_{\rm P}$ can be obtained as
\begin{equation}
V_{\rm eff}(q)
= \frac{3}{2}\frac {U^2\chi_{-}(q)}{1+U\chi_{-}(q)}
+ \frac{1}{2}\frac{U^2\chi_{+}(q)}{1-U\chi_{+}(q)}
- U^2\chi(q)
\label{Eff-Pot}
\end{equation}
\begin{equation}
V_{\rm P}(q)
= \frac{3}{2}\frac {U^2\chi_{-}(q)}{1+U\chi_{-}(q)}
- \frac{1}{2}\frac{U^2\chi_{+}(q)}{1-U\chi_{+}(q)}
+ U^2\chi_1(q)-U
\label {p-poten-s}
\end{equation}
with $\chi_{\pm}(q) = \chi(q)\pm \chi_1(q)$, and 
\begin{equation}
\chi_1(q)=-\frac{T}{N}\sum_k
G_1(k + q) G_1(k). \label{a-Polar-Function}
\end{equation}
The expression for $\chi(q)$ is the same as Eq. (\ref{Polar Function}) where the Green's function $G(k)$ is understood as $G_{11}(k)$.  A simple derivation of these results is presented in Appendix A. 

Following the similar analysis as in the previous subsection, one can take a corresponding approximation for the pairing fluctuation. In the superconducting case, however, besides the diagonal pair propagating (pairing of particles or holes), attention must be paid to the off diagonal pair propagating as well. The latter is the process that the initial state is a pair of particles (holes), while the final state is a pair of holes (particles). Therefore, the pair propagators satisfy a matrix equation. The ladder-diagram approximation with $d$-wave channel interaction is given in Appendix A. The function $P(q)$ appeared in $\Sigma_{11}$ represents the pair propagating. It can be divided into two parts, $P(q) = P_0(q) + P_3(q)$. Their expressions are given by
\begin{eqnarray}
P_0(q)&=&v[D(q)-1-v\Pi_0(q)]/D(q)-v^2\Pi_0(q) \label {P0-eq}\\
P_3(q)&=&v^2\Pi_3(q)[1-D(q)]/D(q) \label {P3-eq}\\
D(q)&=&[1+v\Pi_{+}(q)][1+v\Pi_{-}(q)]-v^2\Pi^2_{3}(q) \label{D-eq}
\end{eqnarray}
with $\Pi_{\pm}(q)=\Pi_0(q) \pm \Pi_1(q) $, and 
\begin{eqnarray}
\Pi_0(q)&=&\frac{T}{N}\sum_k\phi^2(k) [G_0(k) G_0(k-q)-G_3(k) G_3(k-q)] \label{Pi0}\\
\Pi_1(q)&=&\frac{T}{N}\sum_k\phi^2(k) G_1(k) G_1(k-q)   \label{Pi1}\\
\Pi_3(q)&=&\frac{T}{N}\sum_k\phi^2(k) [G_3(k) G_0(k-q)-G_0(k) G_3(k-q)]. \label{Pi3}
\end{eqnarray}
The eigen-equation for determining the function $\phi(k)$ now is extended to
\begin{equation}
\frac{T}{N}\sum_{k'}V_{\rm P}(k-k')[G^2_3(k')+G^2_1(k')-G^2_0(k')]\phi(k')=\phi(k).  \label{s-eigen eq}
\end{equation}
with the largest eigenvalue being unity. This equation is equivalent to Eq. (\ref{gap-eq}) since $\phi(k)$ differs from $\Sigma_1(k)$ by a normalization constant. It also leads to $1+v\Pi_{-}(0)=0$, and thereby $P(q)$ diverges at $q = 0$, which means the existence of the Goldstone mode. 

In terms of the above functions, the diagonal parts of the self-energy can be expressed as
\begin{eqnarray}
\Sigma_0(k)&=&-\frac{T}{N}\sum_qG_0(k-q)V_{\rm eff}(q) \cr\cr
& &-\frac{T\phi^2(k)}{N}\sum_q [G_0(k-q)P_0(q)-G_3(k-q)P_3(q)], \label {0 self energy}\\
\Sigma_3(k)&=& -\frac{T}{N}\sum_qG_3(k-q)V_{\rm eff}(q) \cr\cr
& &-\frac{T\phi^2(k)}{N}\sum_q [G_0(k-q)P_3(q)-G_3(k-q)P_0(q)]. \label {3 self energy} 
\end{eqnarray}
By using the Pauli component of the Green's function, the expression for Eq. (\ref{delta}) can be simplified as
\begin{equation}
\delta = -\frac{2T}{N}\sum_kG_3(k).\label{s-delta}
\end{equation}
So far, we have all the equations for the superconducting case.

\subsection{Q2D approximation}

The Green's function $\hat G(k)$ and the susceptibilities $\chi(q)$'s and $\Pi(q)$'s are defined in three-dimensional space. Actually, in case of $t_z/t\ll 1$, they very weakly depend on the $z$-component variables. However, the dependence on $q_z$ of function $P(q)$ is delicate. Consider the denominator function $D(q)$ at small $\bf q$ and $Z_m$ = 0. Since $\Pi(q)$'s are even functions of $\bf q$, we have
\begin{equation}
D({\bf q},0)\approx c(q_x^2+q_y^2)+c_zq_z^2 \label{asmp}
\end{equation}
where $c$ and $c_z$ are constants. The $q_z^2$-term in Eq.~(\ref{asmp}) comes from the interlayer electron hopping. The ratio $c_z/c$ is much less than unity. If the $q_z$-dependence in $D(q)$ is ignored, the second summations in Eqs. (\ref{0 self energy}) and (\ref{3 self energy}) will be divergent, which implies there will be no superconductivity in the system at finite temperature. This conclusion is consistent with the MWH theorem.\cite{Mermin} We therefore need to keep the $q_z$-dependence in the denominators of $P(q)$'s at least to the order $q^2_z$.

For illustrating our approximation scheme, we firstly consider the case of $Z_m = 0$. Since $\Pi_3({\bf q},0)=0$, we have $P_3({\bf q},0) = 0$ and
\begin{equation}
P_0 = {v^2\over 2}[{\Pi_{-}\over 1+v\Pi_{-}}+ {\Pi_{+}\over 1+v\Pi_{+}}]  -v^2\Pi_0 \label {P00}
\end{equation}
where the arguments $({\bf q},0)$ have been dropped for brevity. As has been mentioned in last subsection, the first denominator $1+v\Pi_{-}$ vanishes at $q = 0$. Even though the second denominator $1+v\Pi_{+}$ is finite at $T < T_c$, it is small. Especially, it vanishes too at $T = T_c$. Therefore, we expand both of the denominators to the order $q^2_z$,
\begin{equation}
1+v\Pi_{\pm}(q) \approx 1 + v\bar\Pi_{\pm}(q)+c^{\pm}_zq^2_z \label{pm expansion}
\end{equation}
with
\begin{eqnarray}
\bar \Pi_{\pm}(q)&=& \Pi_{\pm}(q)|_{q_z=0} \label{bar pi}\\
c^{\pm}_z &=& \frac{v}{2}\frac{d^2}{dq^2_z}\Pi_{\pm}(q)|_{q=0} \label{pmcz1}
\end{eqnarray}
and use $\bar\Pi_{\pm}(q)$ for $\Pi_{\pm}(q)$ in the nominator in Eq. (\ref{P00}). Note that $c^{\pm}_z$ is defined as the derivation in the right-hand side of Eq. (\ref{pmcz1}) at $q = 0$ since where the $q_z$ dependence is important. To evaluate the constants $c^{\pm}_z$, we need to take the derivative of the Green's functions $G(q-k)$'s with respect to $q_z$ as indicated by Eqs. (\ref{Pi0}) and (\ref{Pi1}). By neglecting the $q_z$-dependence of the self-energy, $G(q-k)$'s thereby depend on $q_z$ only via $\xi_{q-k}$. It is expectable that such an approximation does not change the physical result so much. To the second order of $t_z/t$, we obtain\cite{Yan}
\begin{equation}
c^{\pm}_z=\frac{t_z^2vT}{N}\sum_k\phi^2(k)\{[\frac{\partial}{\partial\xi_k}G_3(k)]^2\mp[\frac{\partial}{\partial\xi_k}G_{1}(k)]^2-[\frac{\partial}{\partial\xi_k}G_0(k)]^2\}, \label{pmcs2}
\end{equation}
with 
\begin{eqnarray}
\frac{\partial}{\partial\xi_k}G_0(k)&\approx& 2G_0(k)G_3(k)\cr\cr
\frac{\partial}{\partial\xi_k}G_1(k)&\approx& 2G_1(k)G_3(k)\cr\cr
\frac{\partial}{\partial\xi_k}G_3(k)&\approx& G^2_0(k)-G^2_1(k)+G^2_3(k).  \label{dGf}
\end{eqnarray}
With such an approximated $P_0({\bf q},0)$, the integral over $q_z$ in Eqs. (\ref{0 self energy}) and (\ref{3 self energy}) at $Z_m = 0$ can be taken immediately by neglecting the $q_z$-dependence in the Green's function. Therefore, instead of $P_0({\bf q},0)$ in Eqs. (\ref{0 self energy}) and (\ref{3 self energy}), we insert in a function defined by
\begin{eqnarray}
P^{\rm eff}_0({\bf q},0)&=&\frac{1}{\pi}\int_0^{\pi}dq_z P_0({\bf q},0) \cr\cr
&=& \frac{v^2}{2}[\bar\Pi_{-}f_{-}+\bar\Pi_{+}f_{+}]-v^2\bar\Pi_0,
\label{bar P0}
\end{eqnarray}
with
\begin{eqnarray}
f_{\pm}= \frac{\gamma^{\pm}(q)}{\pi c^{\pm}_z}{\rm atan}[\pi \gamma^{\pm}(q)] \label{fpm}\\ 
\gamma^{\pm}(q)=[{c^{\pm}_z\over 1+v\bar\Pi_{\pm}(q)}]^{1/2}, 
\end{eqnarray}
where again, in the last line of Eq. (\ref{bar P0}), we have dropped the arguments $({\bf q},0)$ for brevity.

We now consider the situation of $Z_m \ne 0$. $D(q)$ is finite in this case. However, in consistent with the approximation for $P_0({\bf q},0)$, we still keep a small $q^2_z$-term in the denominator $D(q)$. Though this $q_z$-dependence is negligible at high temperature, it is reasonable in case of low temperature. According to the expansion by Eq. (\ref{pm expansion}), we expand $D(q)$ as
\begin{equation}
D(q)= \bar D(q)+\epsilon (q)q_z^2 \label {D expansion}
\end{equation}
with
\begin{equation}
\epsilon (q)= [1 + v\bar\Pi_{+}(q)]c^{-}_z+[1 + v\bar\Pi_{-}(q)]c^{+}_z.
\end{equation}
To the order $q^2_z$, this expansion reduces to the result for the case of $Z_m = 0$. Correspondingly, we can define the functions $P^{\rm eff}_0(q)$ and $P^{\rm eff}_3(q)$ by taking the integral over $q_z$. This procedure is equivalent to replacing $1/D(q)$ in Eqs. (\ref{Pi0})-(\ref{Pi3}) with a function $f(q)$ defined by 
\begin{eqnarray}
f(q)= \frac{\gamma(q)}{\pi \epsilon(q)}{\rm atan}[\pi \gamma(q)] \label{feq}\\ 
\gamma(q)=[{ \epsilon(q)\over D(q)}]^{1/2}. 
\end{eqnarray}
With the functions $P(q)$'s in Eqs. (\ref{0 self energy}) and (\ref{3 self energy}) replaced with $P^{\rm eff}(q)$'s, the problem of numerically solving the integral equations is then reduced to a two-dimensional one. All discussed above are for the case of superconducting state. The results for the normal state can be obtained by setting $G_1(k) = 0$. 

\section{Numerical results}

Since the functions $G(k)$, $V_{\rm eff}(q)$ and $P^{\rm eff}(q)$'s are defined in multi-dimensional space, they require huge memory storage in the numerical computation process. Especially, the function $P^{\rm eff}_0(q)$ is singular at $q=0$. Therefore, to carry out the numerical solution, we need to develop numerical scheme to reduce the amount of computation without losing the accuracy. In Appendix B, we present our scheme for the Matsubara frequency summation. The summation is taken over 57 points, a subset of the frequencies, in a sufficiently large range. The cutoff frequencies are $z_c = (2N_c-1)\pi T$ for the fermions and $Z_c = 2(N_c-1)\pi T$ for the bosons, respectively, with $N_c = 1017$. For the typical temperature $T/t \sim 0.01$ under consideration, this means $2N_c\pi T/t \sim 64$. For calculating the function $\chi(q)$, beyond this range, the summation over the terms of $n > N_c$ is analytically carried out by using the asymptotic formula of the Green's function
\begin{equation}
G({\bf k},z_n) \to 1/z_n.
\label{Gatlargen}
\end{equation}
The error of the summation over the terms of $n > N_c$ is of the order $O(z_c^{-3})$.

The convolutions in the momentum space are carried out with fast Fourier transforms (FFTs) on a 128$\times$128 lattice. For inverse transform of $V_{\rm eff}({\bf q},0)$ and $P^{\rm eff}_0({\bf q},0)$, i.e., from momentum space to real space, we have to pay special care. At low temperature, $V_{\rm eff}({\bf q},0)$ has strong peaks near ${\bf q} = (\pi,\pi)$.\cite{Pao} We therefore use a 256$\times$ 256 mesh in momentum space for the inverse transform. The values of $V_{\rm eff}({\bf q},0)$ for this mesh are obtained by local quadratic polynomial interpolation of the smooth functions $\chi(q)$'s given on a 128$\times$128 mesh. On the other hand, the function $P^{\rm eff}_0({\bf q},0)$ has divergently sharp peak at ${\bf q}=0$ and $T\leq T_c$. In Appendix C, we deal with the inverse transform of this function.

The difficulty in solving the eigenvalue problem given by Eq. (\ref{eigen eq}) is that the memory requirement for the coefficient matrix is huge. It is impossible to solve this equation in momentum space. In Appendix D, we rewrite the eigen-equation in real space. At high temperature not too close to $T_c$, Eq. (\ref{eigen eq}) can be solved in real space with a small number of lattice sites in a reduced region. This reduces greatly the amount of numerical calculation work.

The integral equations determining the Green's functions are numerically solved by iteration method. Once a solution at temperature $T$ is obtained, it is then used as an initial input for the next calculation at temperature $T+\delta T$. A more efficient way is to use an extrapolation from the known solutions at temperatures $T_1$ and $T_2$ as the initial input for the next solution at $T_2+\delta T$.

In the present calculation, we set $U/t = 5$ and $t_z/t = 0.01$. All the results presented in the figures are for these parameters. 

\subsection{Eigenvalue $\lambda_d$}

In Fig. 4, we show the result for the eigenvalue $\lambda_d$ as function of $T$ at $\delta = 0.125$. The S-FLEX result is also presented for comparison. Due to the pairing fluctuation, the eigenvalue by the present SP-FLEX approximation is considerably reduced from that of the S-FLEX, giving rise to a lower transition temperature. Moreover, there is a distinguishable difference between their behaviors at temperatures close to $T_c$. By the S-FLEX approximation, we have $d\lambda_d(T)/dT \ne 0$ at $T=T_c$. It means that at $T < T_c$, by keeping no superconducting pairing, the S-FLEX approximation allows a solution of $\lambda_d > 1$. In contrast to this feature of the S-FLEX approximation, the solution for $\lambda_d$ by the SP-FLEX approximation can never get across the line $\lambda_d = 1$. At $T \to T_c$, the pairing fluctuation effect is more pronouncedly with $\lambda_d \to 1$, which conversely suppresses $\lambda_d$. As a result, the curve $\lambda_d(T)$ is smoothly connected to the straight line $\lambda_d = 1$. That is
\begin{equation}
\frac{d}{dT}\lambda_d(T)|_{T_c} = 0. \label{dlambda}
\end{equation}
The inset of Fig. 4 shows that $\sqrt{1-\lambda_d}$ varies nearly linearly as $T\to T_c$, which means $1-\lambda_d \propto (T-T_c)^2$.

A problem then comes in the determination of $T_c$ by $\lambda(T_c)=1$ from the side of normal state. Because of $dT/d\lambda_d = \infty$ at $T = T_c$, a small numerical error in $\lambda_d$ may results in considerable error in $T_c$. Therefore, $T_c$ cannot be accurately determined by the function $\lambda_d(T)$. In our numerical calculations, we solved Eq. (\ref{eigen eq}) using two sets of numbers, 25 and 49 respectively, of the lattice sites in the reduced region (see Appendix D). The solid circles in Fig. 4 represent the numerical results of 49 lattice sites, while the circles are the ones of 25 lattice sites. Close to $T_c$, the difference between the two results are visible. Even with 49 lattice sites, the transition temperature so determined is not reliable.

\vskip -2mm
\begin{figure}
\centerline{\epsfig{file=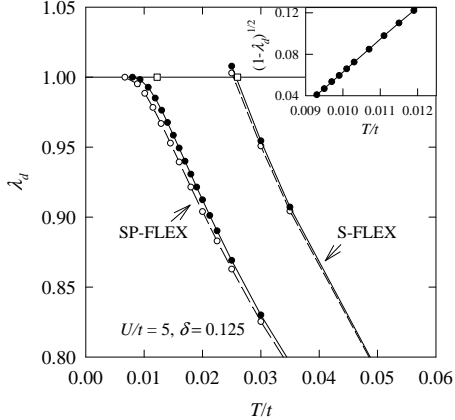,width=7.5 cm}}
\caption{Eigenvalue $\lambda_d$ as function of $T$ at $U/t = 5$ and $\delta=0.125$. SP-FLEX is the present approximation. The result of S-FLEX is also depicted for comparison. The solid circles represent the numerical solution to Eq. (\ref{eigen eq}) solved with 49 lattice sites in a reduced region (see Appendix D). The circles are results of 25 sites. The squares are the transition points obtained by Eq. (\ref{s-eigen eq}). The inset shows $(1-\lambda_d)^{1/2}$ of the SP-FLEX approximation as function of $T$.}
\end{figure}

The problem can be resolved from the superconducting side. Instead of Eq. (\ref{eigen eq}), we solve Eq. (\ref{s-eigen eq}). Since the eigenvalue $\lambda_d = 1$ is known at $T \leq T_c$, Eq. (\ref{s-eigen eq}) can be solved via iteration on the whole lattice. The squares in Fig. 4 denote $T_c$'s obtained from the superconducting side for SP-FLEX and S-FLEX, respectively. The transition temperatures so determined should be reliable.

\subsection{Order parameter}

At low temperatures, we have obtained self-consistent solutions in which $\Sigma_1(k)$ is finite. The symmetry of pairing is $d$-wave, $\Sigma_1({\bf k_x, k_y},z_n)=-\Sigma_1({\bf k_y, k_x},z_n)$. Here, we define the order parameter,
\begin{equation}
\Delta = \Sigma_1({\bf X},z_1)/Z({\bf X},z_1) \label{order-parameter}
\end{equation}
with
\begin{equation}
Z({\bf k},z_n) = 1-\Sigma_0({\bf k},z_n)/z_n 
\end{equation}
and ${\bf X} = (\pi,0)$. The quantity $\Delta$ is a measure of the superconducting gap at the Fermi surface near the point ${\bf X}$.\cite{Pao,Monthoux} Fig. 5 shows the order parameter as function of $T$ at various doping concentrations. The symbols represent the numerical data, while the lines are the extrapolations. As seen from Fig. 5, close to $T_c$, $\Delta$ decreases dramatically. (In the numerical calculation, because of this rapid decreasing, to get an iteration converged at temperature $T+\delta T$ with an initial input extrapolated from some other solutions at and close to temperature $T$, the change $\delta T$ must be very small. At low doping concentrations, close to $T_c$, a change of $\delta T/T \sim 10^{-4}$ at most was allowable in the present calculation.) At $T = T_c$, we have $d\Delta(T)/dT = \infty$. Because of this divergence, $T_c$ can be accurately determined by $\Delta = 0$. 

\vskip -3mm
\begin{figure}
\centerline{\epsfig{file=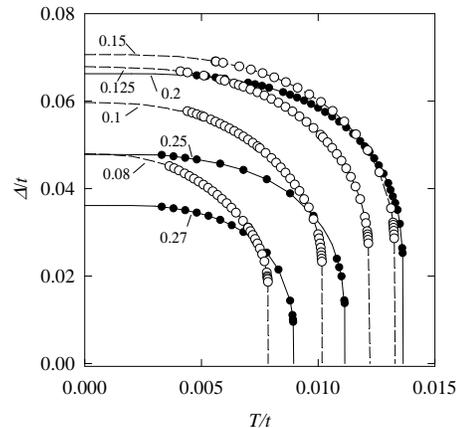,width=7.5 cm}}
\caption{Order parameter $\Delta$ as function of temperature $T$ at various hole concentrations $\delta$. The symbols represent numerical data. The lines are extrapolations.}
\end{figure}

\vskip -3mm
\begin{figure}
\centerline{\epsfig{file=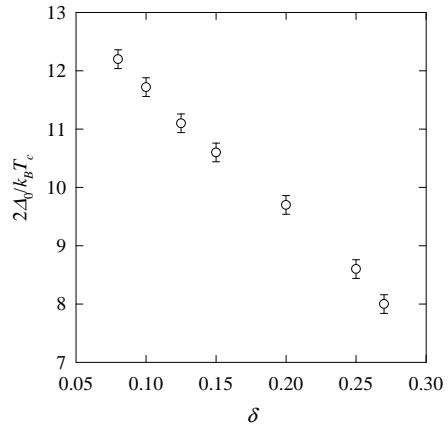,width=7.5 cm}}
\caption{Ratio $2\Delta_0/k_BT_c$ at various hole concentrations $\delta$.}
\end{figure}

On the other hand, at low temperature, $\Delta(T)$ should be flat. We then can obtain the value $\Delta_0 \equiv\Delta(0)$ by extrapolation. In Fig. 6, we show the ratio $2\Delta_0/k_BT_c$ at various doping concentrations. At the underdoping regime, the ratio is about order 10. It decreases with the doping concentration. This ratio is considered as a characterization of the coupling strength of the system. The low doping regime corresponds to strong coupling. With increasing the doping concentration, the coupling gets becoming weak. 

\subsection{Phase diagram}

As mentioned in the previous subsection, the transition temperature $T_c$ is determined by $\Delta(T_c)=0$, which gives accrate solution for $T_c$. The present SP-FLEX result at $U/t=5$ (solid circles) for the boundary of superconducting phase in the $T_c-\delta$ phase diagram is depicted in Fig. 7. All the solid circles are obtained by numerical solutions. The S-FLEX result (circles) is also shown for comparison. Clearly, due to the pairing fluctuation, $T_c$ is considerably reduced from that of the S-FLEX approximation. The reduction is more significant at lower hole doping where $T_c$ decreases with decreasing $\delta$. On the other hand, at large $\delta$, the pairing fluctuation is less pronouncedly. This is consistent with the previous conclusion.\cite{Yan} The results of the previous calculation on the phenomenological model (with coupling constant $J/t=0.2$)\cite{Yan} and the experiments\cite{Tallon} for the cuprate high-temperature superconductors are exhibited in the inset of Fig. 7 for comparison. The behavior of the previous result at small hole doping clearly differs from the experiment and the present calculation. It may stem from the crude treatment of the short-range antiferromagnetic coupling by the phenomenological model. In contrast to the previous result, the present calculation gives a reasonable description of the experiment at small hole doping. 

The dashed line in Fig. 7 is an extrapolation of the numerical result. Unlike the previous case for the phenomenological model,\cite{Yan} the extrapolation gives a nonzero minimum hole concentration, $\delta_m$ very close to 0.05, of the phase boundary. Again, this seems reasonably reflecting the feature of the experimental result. The largest transition temperature $T_{c, \rm max}$ obtained by SP-FLEX is about $0.0137 t$ at $\delta = 0.175$. Using $t \approx 0.6$ eV,\cite{Levin,Yan} we have $T_{c, \rm max}\approx 95$ K. 

\begin{figure}
\centerline{\epsfig{file=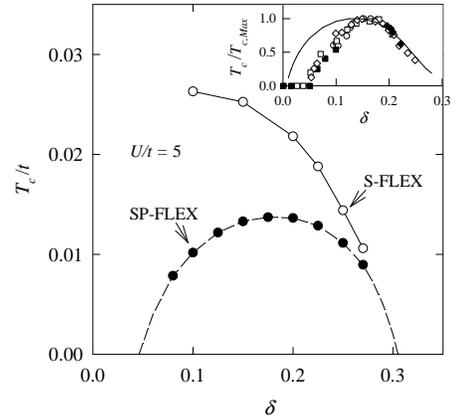,width=7.5 cm}}
\caption{Transition temperature $T_c$ as function of hole concentration $\delta$. The solid-circles and circles are obtained at $U/t =5$ by the SP-FLEX and S-FLEX calculations, respectively. The dashed line is an extrapolation of the numerical result. In the inset, the symbols denote the experimental results for the cuprate high-temperature superconductors,\cite{Tallon} while the solid line is the previous result\cite{Yan} for the phenomenological model.}
\end{figure}

To see why the pairing fluctuation results in reduction of $T_c$, we analyze the self-energy given by Eq. (\ref{self energy}) [which is the same as that given by Eqs. (\ref{0 self energy}) and (\ref{3 self energy}) at $T = T_c$] with $P(q)$ replaced by $P^{\rm eff}(q)$. At $Z_m \simeq 0$, since $V_{\rm eff}(q)$ and $P^{\rm eff}(q)$ have strong negative peaks respectively at ${\bf q \approx Q}\equiv (\pi,\pi)$ and ${\bf q} = 0$, we here make a crude approximation (for the sake of illustration) for the self-energy,
\begin{equation}
\begin{array}{rl}
\Sigma(k)\approx &-G({\bf k-Q},z_n)[{T\over N}\sum\limits_{q\sim Q}V_{\rm eff}(q)]\\\\
&+G({\bf k},-z_n)[{T\phi^2(k)\over N}\sum\limits_{q\sim 0}P^{\rm eff}(q)].
\end{array}\label{Approx self-energy}
\end{equation}
The summations in the square brackets give rise to two negative quantities. At small $\delta$, $\mu \approx 0$, we have $G({\bf k-Q},z_n) \approx -G({\bf k},-z_n)$. Taking into account of these facts, we get $\Sigma(k)\approx -G({\bf k},-z_n)\Gamma_k^2$ with $\Gamma_k^2= a_1 +a_2\phi^2(k)$, where $a_1$ and $a_2$ are two positive constants. Substituting the result into Eq. (\ref{Green's function}), one obtains
\begin{equation}
G({\bf k},\>z_n)\approx{z_n+\xi_k\over 2\Gamma^2_k}
(1-\sqrt{1+{4\Gamma^2_k\over \xi_k^2-z_n^2}}).
\label{Approx Green's function}
\end{equation}
Applying these results to Eq. (\ref{eigen eq}), we see that the factor $G(k)G(-k)$ is reduced, and so is $T_c$. 

Physically, the quantity 
$$a_2=-{T\over N}\sum\limits_{q\sim 0}P^{\rm eff}(q)$$
is a measure of the density of pairs at their excited states. Since the function $P^{\rm eff}(q)$ is given in terms of $\bar\Pi(q)$, we analyze $\bar\Pi(q)=\Pi(q)|_{q_z=0}$ especially at $q=0$. From Eq. (\ref{P Polar Function}), we see that the quantity $\Pi(0)$ comes predominately from the summation over the points close to the Fermi surface in momentum space. At smaller $\delta$, the Fermi surface of the Hubbard model is larger, so is the number of the pairs at their fluctuating states. Therefore, the reduction on $T_c$ is larger at smaller $\delta$.

The form of the Green's function given by Eq. (\ref{Approx Green's function}) implies that there exists a pseudogap in the energy spectrum of the electrons.\cite{Yan,Deisz} Consider the spectral function
$$
\begin{array}{rl}
A({\bf k},E)=&-{1\over \pi}{\rm Im}G({\bf k},E+i0^{+})\\\\
=& {E+\xi_k\over 2\pi \Gamma^2_k}\sqrt{{\xi^2_k+4\Gamma^2_k-E^2\over E^2-\xi^2_k}},
\end{array}
$$
which is nonzero only for $E^2-4\Gamma^2_k<\xi^2_k<E^2$. The noninteraction delta-function peak becomes a square root singularity. Because of the constraint, the area of the ${\bf k}$-space of $A({\bf k},E)\neq 0$ decreases at $E\to 0$, resulting in a suppression of the density of states. Especially, the density of states vanishes at $E=0$ (for $\mu = 0$) since the area becomes to zero and the singularity disappears there. 

On observing the function $\Gamma^2_k$, we note that the pseudogap stems from the spin and pairing fluctuations. Even at $T=0$, the pseudogap remains in the superconducting state because the spin fluctuation (owing to which the superconductivity takes place) and the pairing fluctuation (coming from the Goldstone mode\cite{Yan}) exist. This may explain the recent experiment.\cite{Tallon1}

\subsection{Density of states}

The density of states is defined by
\begin{equation}
\rho(E)=-{1\over \pi N}\sum_{\bf k}{\rm Im}G_{11}({\bf k},E+i0^{+}). \label{DOS}
\end{equation}
The Green's function needs to be analytically continued from the imaginary Matsubara frequency to the real frequency. In terms of an effective self-energy $\tilde\Sigma(k)$ defined by
\begin{equation}
\tilde\Sigma(k)=\Sigma_0(k)+\Sigma_3(k)+\frac{\Sigma^2_1(k)}{z_n-\xi_{\bf k}-\Sigma_0(k)+\Sigma_3(k)},
\end{equation}
the Green's function $G_{11}(k)$ is written as 
\begin{equation}
G_{11}(k)=\frac{1}{z_n-\xi_{\bf k}-\tilde\Sigma(k)}.
\end{equation}
Using Pad\`{e} approximation,\cite{Vidberg} we have obtained the analytical continuation for the effective self energy $\tilde\Sigma(k)$.   

\vskip -3mm
\begin{figure}
\centerline{\epsfig{file=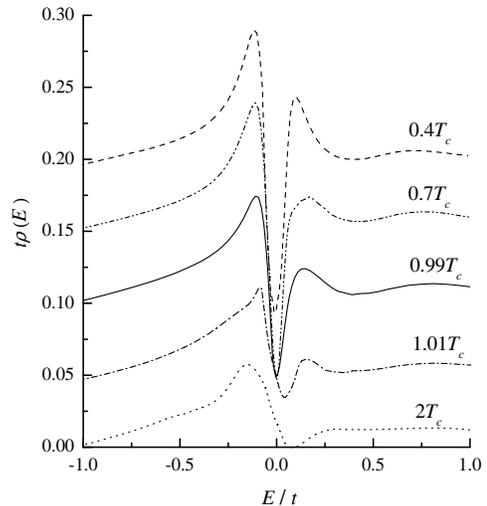,width=7.5 cm}}
\caption{Density of states $\rho(E)$ at $\delta = 0.125$ at various temperatures. For clarity, the $y$-axes for the results of 0.7$T_c$ and 0.4$T_c$ have been offset upwards 0.05 and 0.1, respectively. For 1.01$T_c$ and 2$T_c$, the offsets are downwards 0.05 and 0.1, respectively.}
\end{figure}

The results for the density of state at $\delta = 0.125$ at various temperatures are depicted in Fig. 8. At low temperature, the width of the gap is nearly constant. Below $T_c$, a peak-dip-hump (PDH) structure is clearly seen above the Fermi energy. The positions of the peak and hump are about $\Delta$ and 3$\Delta$, respectively. Such a phenomenon has been observed in the cuprates experiments and has been explained by model calculations.\cite{Shen,DeWilde} The PDH stems from a coupling between the electrons and a collective mode of energy about 2$\Delta$. Below $T_c$, the superconducting gap opens with the maximum of $2\Delta$ appeared in the region near the points $(\pm \pi,0)$ and $(0,\pm \pi)$ in the Brillouin zone. The collective spin fluctuation mode of energy $2\Delta$ and momentum ${\bf Q}$ can then exist in the system. An electron of energy 3$\Delta$ can transit to a state of energy $\Delta$ (at which the DOS has a peak) by exciting the collective mode and losing energy 2$\Delta$. Effectively, the lifetime of the electrons of energy 3$\Delta$ is short, and thereby the dip appears in the DOS. In the FLEX scheme, such a collective mode is described by the effective interaction $V_{\rm eff}$ that is self-consistently determined by the present calculation. 

On the other hand, above $T_c$, there still remains a pseudogap in the DOS. With increasing the temperature, the minimum moves to high energy. As stated in last subsection, the pseudogap comes from both of the spin and pairing fluctuations. At higher temperature, the pairing fluctuation is less important. The electron states ${\bf k}$ and ${\bf k +Q}$ couple with each other mainly through the spin fluctuations. Especially, the degeneracy of those states at the magnetic zone surface is lifted and the weights shift away, resulting in a reduction in the DOS at the corresponding energy.

\subsection{London penetration depth}

In this subsection, we study the magnetic penetration depth. The London penetration depth $\lambda_L$ in $x$-direction is given via 
\begin{equation}
\lambda^{-2}_L=\frac{4\pi ne^2}{m^{*}c^2}+\frac{4\pi}{c^2}C(q)|_{q=0} \label{LC}
\end{equation}
with $C(q)$ the $y$-component current-current correlation function defined by 
\begin{equation}
C({\bf q},\tau-\tau')= -<T_{\tau}[J_y({\bf q},\tau)J_y(-{\bf q},\tau')]>,
\end{equation}
where $<\cdots>$ means a statistical average, $T_{\tau}$ is the imaginary time-$\tau$ ordering operator, and $J_y$
is the $y$-component current operator, 
$${\bf J}({\bf q})=-e\sum\limits_{{\bf k}\alpha}\nabla_{\bf k}\xi_{\bf k}c^{\dagger}_{{\bf k}-{\bf q}/2\alpha}c_{{\bf k}+{\bf q}/2\alpha}.$$
In Eq. (\ref{LC}), $n$ and $m^{*}$ are the number density and the effective mass of electrons, respectively. By using the Ward identity for the current vertex, the quantity $C_0 = C(q)|_{q=0}$ can be written as
\begin{equation}
C_0= \frac{e^2T}{N}\sum_k[G^2_{11}(k)+G^2_{12}(k)]\nabla_{\bf k}\xi_{\bf k}\cdot{\bf v}_k. \label{cccf0}
\end{equation}
with ${\bf v}_k = \nabla_{\bf k}[\xi_{\bf k}+\Sigma_{11}(k)]$. By noting the following equations,
$$G^2_{11}(k)\nabla_{\bf k}[\xi_{\bf k}+\tilde\Sigma(k)]=\nabla_{\bf k}G_{11}(k),$$
$$\frac{T}{N}\sum\limits_k\nabla_{\bf k}\xi_{\bf k}\cdot\nabla_{\bf k}G_{11}(k)=-\frac{n}{m^{*}},$$
we get the expression for $\lambda^{-2}_L$,
\begin{equation}
\lambda^{-2}_L= \frac{4\pi e^2T}{c^2N}\sum_k[G^2_{12}(k){\bf v}_k-G^2_{11}(k){\bf u}_k]\cdot\nabla_{\bf k}\xi_{\bf k}, 
\label{LCF}
\end{equation}
with 
$${\bf u}_k = \nabla_{\bf k}\frac{\Sigma^2_1(k)}{z_n-\xi_{\bf k}-\Sigma_0(k)+\Sigma_3(k)}.$$
By Eq. (\ref{LCF}), $\lambda^{-2}_L$ vanishes identically at and above $T_c$. 

\vskip -3mm
\begin{figure}
\centerline{\epsfig{file=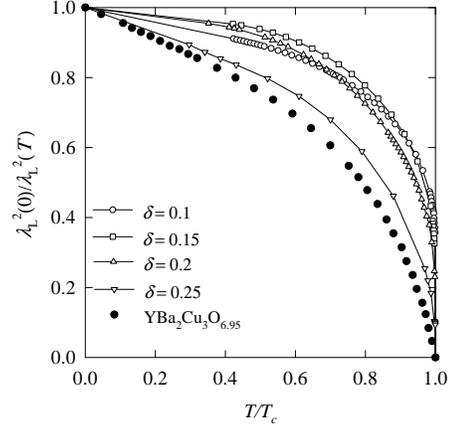,width=7.5 cm}}
\caption{Quantity $\lambda^2_L(0)/\lambda^2_L(T)$ as function of temperature $T$ at various doping concentrations $\delta$. The solid circles denote the experimental results for $a$-axis penetration depth of ${\rm YBa_2Cu_3O_{6.95}}$.}
\end{figure}

Shown in Fig. 9 are the results for the quantity $\lambda^{-2}_L(T)$ as function of temperature $T$ at various doping concentrations. The experimental result for the $a$-axis penetration depth of ${\rm YBa_2Cu_3O_{6.95}}$ is also presented for comparison.\cite{Hosseini} It is well known that $\lambda^{-2}_L(T)$ under $d$-wave pairing symmetry varies linearly with $T$ at low $T$. Using this property, we infer the value $\lambda^{-2}_L(0)$ from extrapolation of the known results at finite temperatures. We then get the zero-temperature superfluid density $n_s$. Figure 10 shows the relationship between $T_c$ and $n_s$ at a number of doping concentrations. This result resembles the experimental observation by Uemura {\it et al.}.\cite{Uemura} They have found a universal linear relation between $T_c$ and $n_s/m^{*}$ at underdoping regime. This behavior cannot be explained by the BCS theory, nor by the S-FLEX scheme in which $T_c$ and $n_s$ do not decrease with $\delta$ decreasing. 

In the mean-field theory, the electron pairs are considered as all in the Bose-Einstein condensate below $T_c$. Since the pairing interaction (stemming from the spin-fluctuation exchange) is stronger at smaller doping concentration, more electron pairs are produced in the condensate. This leads to higher $T_c$ and larger superfluid density. In contrast to the mean-field theory, in the SP-FLEX scheme, the electron pairs are allowed to occupy their excited states. At low temperature, those collective modes are the most available excited states. Even at the ground state, there remains the zero-point motion for the collective modes. So, only a part of the pairs stay in the condensate. As stated earlier, the pairing fluctuation is stronger at smaller $\delta$, resulting in lower $T_c$ and lower $n_s$.

\vskip -3mm
\begin{figure}
\centerline{\epsfig{file=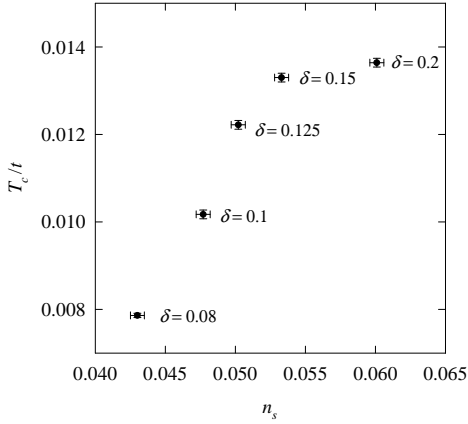,width=7.5 cm}}
\caption{$T_c$ vs. $n_s$ at various doping concentrations $\delta$. }
\end{figure}

\section{Summary}

In summary, we have investigated the $d$-wave superconductivity in the quasi-two-dimensional repulsive Hubbard model. Both of the spin and pairing fluctuations are taken into account in the self-energy. We have self-consistently solved the integral equations for the Green's function. The present calculation reflects a number of features of the experiment results for the cuprate high-temperature superconductors. The calculated boundary of superconducting phase shows a parabolic-like shape, reasonably describing the experiments. The peak-dip-hump structure in the density of states is naturally reproduced by the present calculation. In addition, the present calculation gives reasonable explanations for the temperature dependence of the penetration depth and the relationship between $T_c$ and the superfluid density.

The pairing fluctuation implies that an amount of pairs occupy their excited states. This fluctuation effect leads to the reduction of the condensation of the pairs and thereby the transition temperature. On the other hand, the spin fluctuation produces the attraction between the electrons, resulting in the electron pairing and condensation. In the meanwhile, the spin and pairing fluctuations commonly bring about a pseudogap to the electrons whether the temperature is below or above $T_c$.

\vskip 3mm
\centerline {\bf ACKNOWLEDGMENTS}
\vskip 3mm

The author thanks Prof. C. S. Ting for useful discussions on the related problems. This work was supported by Natural Science Foundation of China under grant number 10174092 and by Department of Science and Technology of China under grant number G1999064509.

\vskip 3mm
\centerline {\bf APPENDIX A}
\vskip 3mm

Considering the convenience for readers, we here present simple derivations of the effective interaction $V_{\rm eff}$, the pairing potential $V_{\rm P}$ and the ladder-diagram approximation for the pair propagators for the superconducting state. 

Firstly, we consider the extension of Fig. 1(b). The interaction $V_{\rm b}(q)$ between $\alpha$-spin electrons can be written as
$$V_{\rm b}(q) = U^2\tilde\chi^{\beta\beta}(q)  \eqno(A1)$$
where $\tilde\chi^{\beta\beta}(q)$ is the density-density response function between the opposite $\beta$-spin electrons. Generally, the function $\tilde\chi^{\alpha\alpha'}$ in the imaginary time-$\tau$ space is defined as
$$\tilde\chi^{\alpha\alpha'}({\bf q},\tau-\tau') = -<T_{\tau}n_{\alpha}({\bf q},\tau)n_{\alpha'}({-\bf q},\tau')>/N \eqno(A2)$$
where $n_{\alpha}({\bf q},\tau)$ is the density operator of $\alpha$-spin electrons. In the Mutsubara-frequency space, the Dyson equation for $\tilde\chi^{\alpha\alpha'}(q)$ reads
$$\tilde\chi^{\alpha\alpha'}(q) = \chi^{\alpha\alpha'}(q)+\sum_{\gamma}\chi^{\alpha\gamma}(q)U\tilde\chi^{-\gamma\alpha'}(q)\eqno(A3)$$ 
where the $\gamma$-summation runs over the up and down spins, and $\chi$'s are the irreducible response functions. Under the ring-approximation, from Eq. (A2), $\chi^{\alpha\alpha}(q)$ is obtained as the same as given by Eq. (\ref{Polar Function}) where the Green's function $G(k)$ is understood as $G_{11}(k)$. $\chi^{\beta\beta}(q)$ is equal to $\chi^{\alpha\alpha}(q)$ because of the equality of the up and down spin electrons. For $\alpha \ne \beta$, we have $\chi^{\alpha\beta}(q)= \chi^{\beta\alpha}(q) \equiv \chi_1(q)$ as given by Eq. (\ref{a-Polar-Function}). Solving Eq. (A3) and substituting the result into Eq. (A1), we have 
$$V_{\rm b}(q) = \frac{U}{2}[\frac{1}{1-U\chi_{+}(q)}-\frac{1}{1+U\chi_{-}(q)}], \eqno(A4)$$
with $\chi_{\pm}(q)=\chi(q)\pm\chi_1(q)$.

Secondly, for the interaction between transverse spins corresponding to Fig. 1(c), we write
$$V_{\rm c}(q) = U^2\tilde\chi^{+-}(q)  \eqno(A5)$$
with $\tilde\chi^{+-}(q)$ the response function between transverse spins defined by
$$\tilde\chi^{+-}({\bf q},\tau-\tau') = -<T_{\tau}S^{+}({\bf q},\tau)S({\bf q},\tau')>/N \eqno(A6)$$
and $S^{+}({\bf q},\tau)=\sum\limits_{\bf k}c^{\dagger}_{{\bf k}\uparrow}(\tau)c_{{\bf k}+{\bf q}\downarrow}(\tau)$. From Eq. (A6), the irreducible function can be obtained as $\chi^{+-}(q) = \chi_{-}(q)$. Since $\tilde\chi^{+-}(q)$ satisfies the Dyson equation $\tilde\chi^{+-}(q) = \chi_{-}(q)-\chi_{-}(q)U\tilde\chi^{+-}(q)$, we have  
$$V_{\rm c}(q) = \frac{U^2\chi_{-}(q)}{1+U\chi_{-}(q)}. \eqno(A7)$$
With the above results, for the effective interaction $V_{\rm eff}(q)=V_{\rm b}(q)+V_{\rm c}(q)-U^2\chi(q)$, we get the expression as given by Eq. (\ref{Eff-Pot}).

Analogously, the interaction corresponding to Fig. 1(f) can be written as 
$$V_{\rm f}(q) = U+U^2\tilde\chi^{\alpha\beta}(q).  \eqno(A8)$$
From Eq. (A3), we have
$$\tilde\chi^{\alpha\beta}(q) = \frac{1}{2}[\frac{\chi_{+}(q)}{1-U\chi_{+}(q)}-\frac{\chi_{-}(q)}{1+U\chi_{-}(q)}].
\eqno(A9)$$
The result for the pairing potential $V_{\rm P}(q)= V_{\rm c}(q)-V_{\rm f}(q)+U^2\chi_1(q)$ is then obtained as given by Eq. (\ref{p-poten-s}). Again, the term $U^2\chi_1(q)$ eliminates a second-order double counting.

Finally, the function $P(q)$ appeared in the element $\Sigma_{11}(k)$ of the self-energy can be expressed as 
$$P(q) = v^2[\tilde\Pi_{11}(q)-\Pi_{11}(q)] \eqno(A10)$$
where $\tilde\Pi_{11}(q)$ is the pair-pair response function, and $\Pi_{11}(q)$ is the irreducible part (or the pair susceptibility, which eliminates the second-order double counting). The function $\tilde\Pi_{\mu\nu}({\bf q},\tau-\tau')$ is defined as  
$$\tilde\Pi_{\mu\nu}({\bf q},\tau-\tau') = -<T_{\tau}p_{\mu}({\bf q},\tau)p^{\dagger}_{\nu}({\bf q},\tau')>/N, \eqno(A11)$$
with $\mu, \nu$ = 1, 2, $p_1({\bf q}) = \sum\limits_{\bf k}\phi(k)c_{{\bf q}-{\bf k}\downarrow}c_{{\bf k}\uparrow}$ and $p_2({\bf q}) = \sum\limits_{\bf k}\phi(k)c^{\dagger}_{{\bf k}-{\bf q}\uparrow}c^{\dagger}_{-{\bf k}\downarrow}$. Strictly speaking, $p_1$ and $p_2$ are not the Schr\"odinger operators since $\phi(k)$ depends on the Matsubara frequency. Here, we formally regard them as only depending on the momentum. At the end, we extend the result to include the frequency dependence. (Alternatively, one can draw the Feyman diagram from the beginning. The final result is the same.) From Eq. (A11), the expressions for the irreducible susceptibilities can be obtained as Eqs. (\ref{Pi0})-(\ref{Pi3}). The Dyson equation for $\tilde\Pi_{\mu\nu}$ in matrix form is
$$\hat{\tilde\Pi}(q)=\hat\Pi(q)-v\hat\Pi(q)\hat{\tilde\Pi}(q). \eqno(A12)$$ 
The diagonal parts of the Pauli components of the matrix $\hat P(q) = v^2[\hat{\tilde\Pi}(q)-\hat\Pi(q)]$ can be expressed as Eqs. (\ref{P0-eq}) and (\ref{P3-eq}).

\vskip 3mm
\centerline {\bf APPENDIX B}
\vskip 3mm
In this appendix, we intend to develop an algorithm for the approximate summation of a series. It is analogous to Simpson's integral method. Firstly, we consider the following summation,
$$S(n_0,n_2)= \sum_{n=n_0}^{n_2}f(n),
\eqno (B1)$$
where $n_2 = n_0+2h$ with $h$ an integer. Suppose $f(x)$ is a smooth function over the range $n_0<x<n_2$. We then expand $f(n)$ as 
$$f(n)\approx f(n_0)+c_1(n-n_0)+c_2(n-n_0)^2
\eqno (B2)$$
where $c_1$ and $c_2$ are constants. With the given values $f_j \equiv f(n_0+jh)$, for $j$ = 0, 1 and 2, the constants can be expressed as
$$
\begin{array}{rl}
c_1=& (-3f_0+4f_1-f_2)/2h \cr\cr
c_2 =& (f_0-2f_1+f_2)/2h^2. 
\end{array}
$$
Substituting Eq. (B2) into Eq. (B1), we get
$$S(n_0,n_2)\approx \frac{h}{6}(2+3y+y^2)(f_0+f_2)+\frac{h}{3}(4-y^2)f_1,
\eqno (B3)$$
with $y = 1/h$. Therefore, the summation over the entire range $[n_0,n_2]$ can be obtained approximately with only three values $f_0$, $f_1$ and $f_2$ given. At $y\to 0$, Eq. (B3) reduces to the Simpson rule. With the approximation (B2), we even can carry out a summation over a part of the range $[n_0,n_2]$. For more general uses, for $n_0\leq m \leq n_2$, we have 
$$S(n_0,m)\approx Af_0 + Bf_1 +Cf_2  \eqno (B4)$$
with
$$
\begin{array}{rl}
A=& h(y+z)[1-3z/4+z(y+2z)/12] \\\\
B=& h(y+z)z[1-(y+2z)/6]       \\\\
C=&-h(y+z)z[1-(y+2z)/3]/4     
\end{array} 
$$
and $z=(m-n_0)/h$. 

Now, we consider the summation $S(1,\infty)$. When $f(n)$ deceases fast at $n\to \infty$, $S(1,\infty)$ can be obtained approximately over a finite range with the cutoff number sufficiently large. We may divide this range into several blocks within each of which $f(x)$ can be regarded as smooth function and thereby the above algorithm can be applied. At most cases, $f(n)$ may vary fast at small $n$. Therefore, the stride $h$ should be shorter at smaller $n$. Here, we introduce a point-selection program. Consider $L$ successively connected blocks. The selected points divide each block into $M-1(\geq 2)$ equal-spaced segments; each block contains $M$ points. The stride (the length of the segment) in the $l$-th block is $h_l = h^{l-1}$ with $h$ a constant integer number. By such a program, the number of the $j$-th point in the $l$-th block is
$$n_{[j,l]}= (j-1+\frac{M-1}{h-1})h^{l-1}-\frac{M-h}{h-1},\eqno(B5)$$
for $j=1,2,\cdots,M$ and $l=1,2,\cdots,L$. The cutoff number is $N_c = n_{[M,L]}$. By repeatedly using the above summation scheme, one can get approximately
$$S(1,\infty)\approx \sum_p w_p f(n_p), \eqno (B6)$$
where $p$ runs over the $L(M-1)+1$ selected points, and $w_p$ is the weight at point $p\equiv[j,l]$; note that because of $n_{[M,l]}= n_{[1,l+1]}$, such of these points should be counted once in the summation. The point-selection program uniquely determines the weights. If $M\geq 5$ is an odd number, applying the above summation scheme, we get

\begin{tabular}{ll}\cr
&$w_{[1,1]}$ = 1\cr\cr
&$w_{[j,l]}$=~$(4h^{l-1}-h^{1-l})/3$,
~~{\rm for}~$j=2,4,\cdots,M-1$\cr\cr 
&$w_{[j,l]}$=~$(2h^{l-1}+h^{1-l})/3$, 
~~{\rm for}~$j=3,5,\cdots,M-2$\cr\cr
&$w_{[M,1]}$=~$(2h+3+h^{-1})/6$ \cr\cr
&$w_{[M,l]}$=~$w_{[l+1,1]}=h^l(1+h^{-1})/3+(1+h)h^{-l}/6$,\cr\cr
&~~~~~~~~~~~~~~~~~~~~~~~~~{\rm for}~$l=2,\cdots,L-1$\cr\cr
&$w_{[M,L]}$=~$h^{L-1}/3+1/2+h^{1-L}/6$.\cr\cr
\end{tabular}

To justify the above summation scheme, we here give an example. Consider the summation
$$S=\sum_{j=1}^{\infty}\frac{2}{4j^2-1}=1. \eqno (B7)$$
Applying the above scheme with $h$ =2, $L$ = 7, and $M$ = 9, we have
$$Sum=\sum_p \frac{2w_p}{4n_p^2-1}= 0.99952. \eqno (B8)$$
The relative error is $(Sum-S)/S = -4.8\times 10^{-4}$. Note that the terms under the summation in Eq. (B7) decreases as $1/2j^2$ at large $j$. If one uses the known result 
$$\sum_{j=1}^{\infty}j^{-2} = \pi^2/6,$$
the accuracy of the summation can be improved much better. Instead of Eq. (B8), we calculate the following summation
$$Sum=\sum_p w_p(\frac{2}{4n_p^2-1}-\frac{1}{2n_p^2})+\pi^2/12. \eqno (B9)$$
With such an arrangement, the value of the brackets in the $p$-summation decreases as $O(n_p^{-4})$ at large $n_p$. This summation gives a very accurate result $Sum = 1.00000015$, with a small relative error $1.5\times 10^{-7}$ only.  
 
In some cases, with the given points $n_{[j,l]}$ selected in advance, we need to calculate the summations, $S(1,n)$, with $n_{[j_0,l_c]}\leq n < n_{[j_0+1,l_c]}$ and $l_c \leq L$. In these cases, because the summation over a range needs three points at least, the terminate number $n_c \equiv n_{[j_c,l_c]}$ is then determined as follows
$$
j_c = \cases{3,&if $j_0=1$\cr
	j_0+1,&otherwise.\cr}
$$
Since the summation $S(1,n)$ can be expressed as $S(1,n) = S(1,n_{[1,l_c]}-1)+ S(n_{[1,l_c]},n)$, the weights at the points $n_p < n_{[1,l_c]}$ as tabulated above is unchanged, while at the points $n_{[j,l_c]}$, for $j = 1,\cdots, j_c$, $w_p$ should be reevaluated according to the scheme as given by Eq. (B4)

For testing the accuracy of the scheme for summations of finite terms, we consider the following example
$$S_n=\sum_{j=1}^n\frac{2}{4j^2-1}=\frac{2n}{2n+1}. \eqno (B10)$$
With the $[h,L,M]=[2,7,9]$ program, the summation is approximated as
$$Sum=\sum_p \frac{2w^n_p}{4n_p^2-1} \eqno (B11)$$
where we use the superscript $n$ indicating the $n$-dependence of the weights. The results for $Sum/s_n$ are given in Table I. The relative error $R.E. = (Sum-S_n)/S_n$ is less than $10^{-4}$.

Finally, we give the expression for the susceptibility $\chi$. Since it is even for $Z_m$, we only consider the case of $m\geq 0$. In real space, it is given by
$$
\begin{array}{rl}
\chi (r,Z_m)=&T\sum\limits_{n=-\infty}^{\infty}G(r,z_n)G(r,z_n+Z_m)\cr\cr
 =&T\{2\sum\limits_{n=1}^{\infty}G(r,z_n)G(r,z_n+Z_m)\cr\cr
&~~~~+2\sum\limits_{n=1}^{[m/2]}G(r,z_n)G(r,z_n-Z_m) \cr\cr
& ~~~~+G(r,z_{\bar n})G(r,-z_{\bar n})|_{{\bar n}=(m+1)/2} ^{\rm ~if~m~is~odd}\} \cr\cr
=&T\{2\sum\limits_{p}w_pG(r,z_{n_p})G(r,z_{n_p}+Z_m)\cr\cr
&~~~~+2\sum\limits_{p}w_p^{[m/2]}G(r,z_{n_p})G(r,z_{n_p}-Z_m) \cr\cr
& ~~~~+G(r,z_{\bar n})G(r,-z_{\bar n})|_{{\bar n}=(m+1)/2} 
^{\rm ~if~m~is~odd}\} \cr\cr
& ~~~~+\delta\chi(r,Z_m) 
\end{array} 
$$
where $[m/2]$ is the integer part of $m/2$, and the last term is given by 
$$\delta\chi (r,Z_m)= 2T\sum\limits_{n=N_c+1}^{\infty}G(r,z_n)G(r,z_n+Z_m). $$
Because of $G(r,z_n)\to \delta_{r0}/z_n$ at $n\to \infty$, we have
$$
\delta\chi (r,Z_m)=\cases {
-\frac{\delta_{r0}}{m\pi^2T}\sum\limits_{n=N_c+1}^{N_c+m}\frac{1}{2n-1},& if $m > 0$\cr\cr
-\frac{2\delta_{r0}}{\pi^2T}[\frac{\pi^2}{8}-\sum\limits_{n=1}^{N_c}\frac{1}{(2n-1)^2}],&if $m=0$. }
$$
Note that $G(r,z_n-Z_m) = G^{\ast}(r,Z_m-z_n)$. Therefore, the Green's function at negative Matsubara frequency can be determined from its complex counterpart at the positive frequency. For a given point-selection program, $G(r,z_{n_p})$ is known. Those values at $Z_m\pm z_n$ can be evaluated by interpolation, or $G(r,z_n+Z_m)\approx \delta_{r0}/(z_n+Z_m)$ if $n+m > N_c$.

\vskip 2mm
\centerline {\bf APPENDIX C}
\vskip 2mm

In this Appendix, we discuss the problem of inverse Fourier transform of $P_0^{\rm eff}({\bf q},0)$. A typical result for $P_0^{\rm eff}({\bf q},0)$ is shown in Fig. 11. This function behaves as $P_0^{\rm eff}({\bf q},0) \propto 1/\sqrt{a^2+q^2}$ (with $q^2 = q_x^2+q_y^2$) at $q \to 0$. The constant $a$ vanishes at $T \leq T_c$. Even at $T > T_c$ but close to $T_c$, $a$ is very small. Physically, it means that the pairing fluctuation is defined in a long range in real space. Especially, at $T\leq T_c$, the range is infinite. Therefore, its primary form is not suitable for numerical inverse Fourier transform on a finite lattice. 

\begin{figure}
\centerline{\epsfig{file=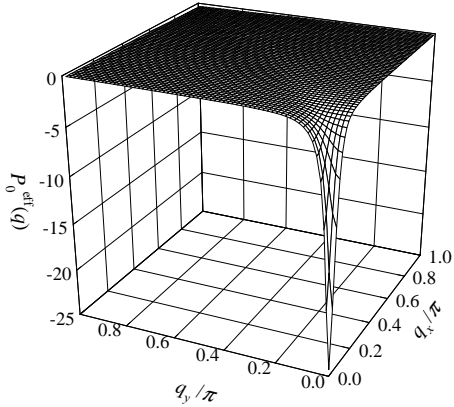,width=8 cm}}
\vskip -8mm
\caption{Function $P_0^{\rm eff}(q)$ in unit of $2t$ at $Z_m =0$, $\delta=0.125$, $U/t=5$, and $T/t=0.0124$.}
\end{figure}

The function $P_0^{\rm eff}({\bf q},0)$ can be divided into the `singular' part $c/\sqrt{a^2+q^2}$ (with $c$ a constant) and the regular one. There is no problem in the inverse Fourier transform for the latter one. For the `singular' part, the task is to calculate the integral
$$F(j_x,j_y)=\int_0^{\pi}dq_x\int_0^{\pi}dq_y\frac{\cos (q_xj_x)\cos (q_yj_y)}{\sqrt{a^2+q^2}}, \eqno(C1)$$
where $j_x$ and $j_y$ are the coordinates of a lattice site. Repeatedly integrating by part, we get
$$
\begin{array}{ll}
F(j_x,j_y)=&-j_y^2\int_0^{\pi}dq_x\int_0^{\pi}dq_yf_1(q)\cos (q_xj_x)\cos (q_yj_y) \\ \\
&+(-1)^{j_y}\int_0^{\pi}dq_xf_2(q_x)\cos (q_xj_x)\\ \\
&-j_x\int_0^{\pi}dq_xf_3(q_x)\sin (q_xj_x)\\ \\
&-(-1)^{j_x} f_3(\pi),
\end{array}
\eqno (C2)$$
with
$$
\begin{array}{ll}
f_1(q)=& q_y\ln(q_y+\sqrt{a^2+q^2})-\sqrt{a^2+q^2} \\ \\
f_2(q_x)=&\ln(\pi+\sqrt{a^2+\pi^2+q_x^2})\\ \\
f_3(q_x)=&q_x(\ln\sqrt{a^2+q_x^2}-1)+a~{\rm atan}(\frac{q_x}{a}).
\end{array}
$$
By this way, all $f$'s are regular functions.

However, because the large factors $j_y^2$ and $j_x$ at long distances, we need to carry out the inverse Fourier transforms in Eq. (C2) with high accuracy. The numerical method of the fast Fourier transforms amounts to applying the trapezoidal rule to the integral in Eq. (C2). One may use a very dense mesh in the Brillouin zone for the transforms. But, this is uneconomical in the present numerical process since such transforms need to be repeatedly performed. In fact, errors in the numerical integration stem mainly from the rapid oscillation behavior in the integrand. Here, we present our scheme for these integrals in Eq. (C2). Essentially, we need to deal with the following integrals,
$$
F_c(j)=\int_0^{\pi}dq f(q)\cos (qj) \eqno(C3)$$
$$
F_s(j)=\int_0^{\pi}dq f(q)\sin (qj) \eqno(C4)
$$
where $f(q)$ is a regular and smooth function over $(0,\pi)$. 

Firstly, we consider the simple case,
$$
I(q_1,q_3)\equiv \int_{q_1}^{q_3}dq f(q)\cos (qj)\eqno (C5)$$
where $(q_1,q_3)$ is a small range. The middle point is $q_2$. Within this range, $f(q)$ can be expressed as
$$f(q)\approx a_1+a_2(q-q_1)+a_3(q-q_1)^2.
\eqno (C6)$$
The constants $a$'s are determined by the values $f_j \equiv f(q_j)$, 
$$
\begin{array}{rl}
a_1=&f_1\\ \\
a_2=& (-3f_1+4f_2-f_3)/2h \cr\cr
a_3 =& (f_1-2f_2+f_3)/2h^2,
\end{array}
$$
where $h=q_2-q_1$. Now, repeatedly integrating Eq. (C5) by part and use Eq. (C6), we obtain
$$
I(q_1,q_3)= C_1\cos (q_1j)+ C_3\cos (q_3j)
+S_1\sin (q_1j)+ S_3\sin (q_3j)
\eqno (C7)$$
with
$$
\begin{array}{rl}
C_1=&(3f_1-4f_2+f_3)h/2x^2\\ \\
C_3=&(f_1-4f_2+3f_3)h/2x^2\\ \\
S_1=&[f_3-2f_2-(x^2-1)f_1]h/x^3\\ \\
S_3=&[(x^2-1)f_3+2f_2-f_1]h/x^3 
\end{array}
$$
and $x=jh$. It is expectable that the result given by Eq. (C7) is more accurate than the trapezoidal rule.

Now, dividing the range $(0,\pi)$ into $2M$ equal spaced pieces, we have
$$
F_c(j)=\sum_{k=1}^MI(q_{2k-1},q_{2k+1}) \eqno(C8)$$
with $q_k = (k-1)h$ and $q_{2M+1}=\pi$. Using the result as given by Eq. (C7), we get
$$
\begin{array}{rl}
F_c(j)=&2w_1(x)[2\sum\limits_{k=2}^Mf_{2k-1}\cos(q_{2k-1}j)+f_1+(-1)^jf_{2M+1}]\\\\
&+4w_2(x)\sum\limits_{k=1}^Mf_{2k}\cos(q_{2k}j)
\end{array}
\eqno(C9)$$
where the functions $w_1$ and $w_2$ are given by
$$
\begin{array}{rl}
w_1(x)=&\frac{h}{4x^2}[3-\frac{2\sin(2x)}{x}+\cos(2x)],
\\\\
w_2(x)=&\frac{h}{x^2}(\frac{\sin x}{x}-\cos x).
\end{array}
$$
Define a new discrete function
$$g_k = (-1)^kf_k, ~~~~{\rm for}~~k=1,\cdots, 2M+1.$$
With this definition, equation (C9) can be rewritten as
$$
F_c(j)=w_1(x)\{C_j[f]-C_j[g]\}+w_2(x)\{C_j[f]+C_j[g]\}
\eqno(C10)$$
where $C_j[f]$ is the cosine Fourier transform of function $f$ defined by
$$
C_j[f]=2\sum\limits_{k=2}^{N-1}f_{k}\cos(q_{k}j)+f_1+(-1)^jf_N$$
with $N=2M+1$. Therefore, the function $F_c(j)$ can be evaluated by the FFT via Eq. (C10).

Similarly, one can get
$$
\begin{array}{rl}
F_s(j)=&w_1(x)\{S_j[f]-S_j[g]\}+w_2(x)\{S_j[f]+S_j[g]\}\\\\
&+{h\over x}[1+{\sin(2x)\over 2x}-{1-\cos(2x)\over x^2}][f_1-(-1)^jf_N]
\end{array}
\eqno(C11)$$
where $S_j[f]$ is the sine Fourier transform of function $f$ defined by
$$
S_j[f]=2\sum\limits_{k=2}^{N-1}f_{k}\sin(q_{k}j).$$

\vskip 4mm
\centerline {\bf APPENDIX D}
\vskip 4mm

In this Appendix, we rewrite the eigen-equation (\ref{eigen eq}) in a form more convenient for the numerical calculation. We intend to solve the equation in real space in order to get rid of the prohibitive storage requirement for the coefficient matrix in momentum space.

We can apply the frequency-summation scheme just developed in Appendix B to the present case, so reducing the memory size.
However, to solve the eigen-equation in momentum space still requires tremendous memory size. In some cases, fortunately, the pairing function is short-ranged in real space. We therefore solve the equation in real space. To transform Eq. (\ref{eigen eq}) into real space, one needs to maintain the matrix of the coefficients to be symmetrical. In the follows, we present the transformation procedure.

(A) Define functions $f({\bf k},n_p)$ and $\psi({\bf k},n_p)$ as
$$f({\bf k},n_p)\equiv\sqrt {Tw_pG({\bf k},z_{n_p})G(-{\bf k},-z_{n_p})}, \eqno(D1)$$
$$\psi({\bf k},n_p)\equiv f({\bf k},n_p)\phi({\bf k},z_{n_p}), \eqno(D2)$$
where $w_p$ is the weight at frequency $z_{n_p}$ as introduced in Appendix B. In real space, Eq. (\ref{eigen eq}) is transformed to
$$\sum_{{\bf r'}p'}A({\bf r},n_p;{\bf r'},n_{p'})\psi({\bf r'},n_{p'})=\lambda\psi({\bf r},n_p) \eqno(D3)$$
with
$$A({\bf r},n_p;{\bf r'},n_{p'})=\sum_{\bf R}f({\bf r-R},n_p)W({\bf R},n_p,n_{p'})f({\bf R-r'},n_{p'}) \eqno(D4)$$
and
$$W({\bf R},n_p,n_{p'})=V_{\rm P}({\bf R},z_{n_p}-z_{n_{p'}})+V_{\rm P}({\bf R},z_{n_p}+z_{n_{p'}}). \eqno(D5)$$

(B) Further more, because of the lattice symmetry, we need only to consider the lattice sites $[r]$ of $0 \leq r_y < r_x$. Those lattice sites of $r_x = r_y$ are excluded since the $d$-wave pairing is under consideration. Define
$$y({\bf r},n)=2\sqrt{2\over d_r}\psi({\bf r},n) \eqno(D6)$$
$$F({\bf r,R},n)={1\over\sqrt{d_rd_R}}\sum_gs_gf(g{\bf r-R},n) \eqno(D7)$$
where $d_r = 1+\delta_{r_y0}$, the $g$-summation runs over the operations of group $C_{4v}$, $s_g = \pm 1$ is the sign factor of the $d$-wave function $\phi({\bf r},z_n)$ under the operation $g$, and $g{\bf r}$ denotes a site coming from ${\bf r}$ operated by $g$. Accordingly, define the new matrix
$$M({\bf r},n_p;{\bf r'},n_{p'})=\sum_{[\bf R]}F({\bf r,R},n_p)W({\bf R},n_p,n_{p'})F({\bf r',R},n_{p'}) \eqno(D8)$$
where again $[{\bf R}]$-summation runs over those lattice sites of $0\leq R_y < R_x$. By so doing, the eigen equation reads,
$$\sum_{[{\bf r'}]p'}M({\bf r},n_p;{\bf r'},n_{p'})y({\bf r'},n_{p'})=\lambda y({\bf r},n_p). \eqno(D9)$$

\begin{figure}
\centerline{\epsfig{file=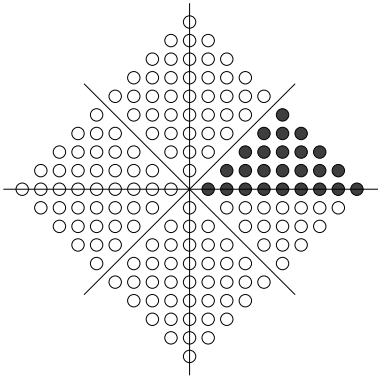,width=5 cm}}
\vskip 2mm
\caption{Sketch of the lattice sites used for solving Eq. (D9). Solid circles show the reduced region.}
\end{figure}

A sketch of the lattice sites is shown in Fig. 12. The reduced region $[r]$ is taken as the shaded sites, which is suitable for the description of $d$-wave pairing. In our numerical calculation, the total number of sites of $[r]$ is $N_r =49$. The normalization condition for $y({\bf r},n_p)$ is
$$\sum_{[{\bf r}]p}y^2({\bf r},n_p) = 1 . \eqno(D10)$$
By this condition, the coupling constant $v$ is given as $v = \lambda /2$. Shown in Fig. 13 are the typical results for $f(r_x,1)$ and $\phi(r_x,z_1)$. Clearly, they are short-ranged functions. In passing, we compare the sizes of matrices of coefficients required for solving the eigen-equation respectively in real and momentum spaces. The dimension of matrix $M$ in Eq. (D8) is $(N_rM_0)\times (N_rM_0) = 2793\times 2793$, where $M_0 =57$ is the number of selected Matsubara frequency. However, in momentum space with a $128\times 128$ mesh, even making use of the lattice symmetry, the dimension is $(N_kM_0)\times (N_kM_0) = 118560\times 118560$, where $N_k = 32\times 65$ is the number of momentum ${\bf k}$ with $0\leq k_y < k_x\leq \pi$.

\begin{figure}
\centerline{\epsfig{file=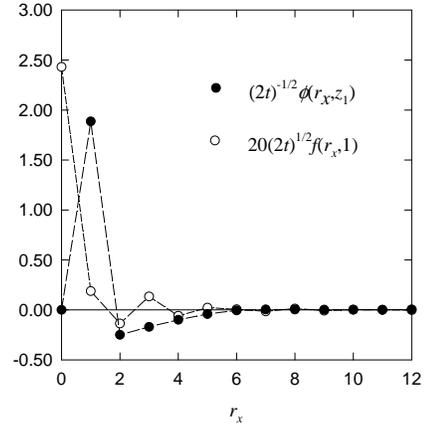,width=7.5 cm}}
\caption{Functions $f(r_x,1)$ and $\phi(r_x,z_1)$ at $r_y = 0$, $\delta=0.125$, $U/t=5$, and $T/t=0.0124$. The dashed lines are for the eyes.}
\end{figure}

\begin{table}
\caption{Ratio of $Sum$ given by Eq. (A11) and $S_n = 2n/(2n+1)$ at various $n$. $R.E.$ represents the relative error.}
\vskip 2mm
\begin{tabular}{rcc}
$n$ & $Sum/S_n$ & $R.E.$ \\ \hline
      5.    &    1.000000    &    0\\
      10.    &    1.000047    &    0.000047\\
      50.    &    1.000012    &    0.000012\\
      100.    &    1.000013    &    0.000013\\
      500.    &    1.000013    &    0.000013
\end{tabular}
\end{table}

\end{document}